\documentclass{ifacconf}

\usepackage{graphicx,xcolor,amsmath,siunitx,amssymb,mathtools}      
\usepackage{natbib,tikz}        

\DeclareMathOperator*{\subject_to}{s. t.\: }

\newtheorem{remark}{Remark}

\usepackage{amsmath,mathtools,tikz,booktabs,bm,multirow,pgfplots}
\usepackage[implicit = false]{hyperref}
\usetikzlibrary{shapes.geometric}
\usepackage{grffile}

\pgfplotsset{compat=newest}
\usetikzlibrary{plotmarks}
\usetikzlibrary{arrows.meta}
\usepgfplotslibrary{patchplots}

\definecolor{mycolor1}{rgb}{0.00000,0.44700,0.74100}%
\definecolor{mycolor2}{rgb}{0.85000,0.32500,0.09800}%
\definecolor{mycolor3}{rgb}{0.46600,0.67400,0.18800}%
\definecolor{dark-gray}{gray}{0.35}
\definecolor{myred}{rgb}{0.6350, 0.0780, 0.1840}
\definecolor{mygreen}{rgb}{0.4660, 0.6740, 0.1880}
\definecolor{myblue}{rgb}{0, 0.4470, 0.7410}

\begin{document}
\begin{frontmatter}

\title{Inference in Latent Force Models Using Optimal State Estimation\thanksref{footnoteinfo}} 

\thanks[footnoteinfo]{This project has received funding from the European Research Council (ERC) under the European Union’s Horizon 2020 research and innovation programme (grant agreement No 948679).
}

\author[IRT]{Tobias M. Wolff} 
\author[IRT]{Victor G. Lopez} 
\author[IRT]{Matthias A. Müller}
\author[Vanderbilt]{Thomas Beckers}

\address[IRT]{Leibniz University Hannover, Institute of Automatic Control, Germany, (e-mail: \{wolff, lopez, mueller\}@irt.uni-hannover.de).}
\address[Vanderbilt]{Department of Computer Science and Department of Mechanical Engineering, Vanderbilt University, Nashville, TN 37212, USA, (e-mail: thomas.beckers@vanderbilt.edu)}

\begin{abstract}                
	Latent force models, a class of hybrid modeling approaches, integrate physical knowledge of system dynamics with a latent force - an unknown, unmeasurable input modeled as a Gaussian process. In this work, we introduce two optimal state estimation frameworks to reconstruct the latent forces and to estimate the states. In contrast to state-of-the-art approaches, the designed estimators enable the consideration of system-inherent constraints. Finally, the performance of the novel frameworks is investigated in several numerical examples. In particular, we demonstrate the performance of the new framework in a real-world biomedical example - the hypothalamic-pituitary-thyroid axis - using hormone measurements. 
\end{abstract}


\end{frontmatter}
\section{Introduction}
\label{sec:introduction}
In many applications, we encounter the problem of reconstructing a latent force that affects a dynamical system only based on output measurements. This problem can be addressed by latent force models \citep{Lawrence2006,Gao2008,Alvarez2013}, which combine models derived from first principles and latent (i.e., unmeasurable) forces. They have, e.g., been successfully applied to predict satellite orbits \citep{Rautalin2017}, to describe electric propagation in deep brain stimulation \citep{Alvarado2014}, and to model the thermal dynamics of buildings \citep{Ghosh2015}. Typically, the derived mathematical model is available in form of an ordinary (or partial) differential equation. The latent force\footnote{Please note that the term \textit{latent force} comes from the machine learning community. In the systems and control field, one might rather use the term \textit{unknown input}.}, an unmeasurable and uncontrollable input to the system, is modeled by a Gaussian process (GP) \citep{Rasmussen2006}. It can represent a disturbance (which is not independent and identically distributed) such as, e.g., the wind that affects a car, plane, or ship \citep{Landgraf2022,Grasshoff2019}. Alternatively, it can represent internal variables of the system like, e.g, a certain chemical concentration that influences the dynamics of some biomedical system, compare the work of \cite{Lawrence2006}. 

In \textit{linear} latent force models (i.e., when the considered model is linear), the ordinary differential equation can be solved analytically in dependence of the latent force. Based on the covariance function of the latent force, one can compute a covariance function of the outputs of the differential equation. Using this ``informed" covariance function, standard GP regression can be performed \citep{Lawrence2006,Rasmussen2006}. 

A different approach to latent force models, which will also be exploited in this work, is to use the state-space representation of GPs \citep{Hartikainen2010}. In this case, one transforms the available differential equation to a state-space model, which is augmented by the state-space representation of GPs \citep{Hartikainen2011,Grasshoff2019}. To perform inference in the linear case, the Kalman filter and the Rauch-Tung-Striebel (RTS) smoother are used \citep{Kalman1960,Rauch1965}. The advantages of this approach are the computational simplicity and that the differential equation does not need to be solved analytically. Furthermore, this approach can easily be extended to nonlinear latent force models (where the physical model is allowed to be nonlinear). One augments the nonlinear state-space model by the state-space representation of GPs and performs inference with nonlinear Kalman filter and RTS smoother extensions \citep{Hartikainen2012,Landgraf2022}. However, since this approach does not allow to include any constraints, the estimates of the latent force can become physically implausible. In that case, the estimates are meaningless and cannot be interpreted. This is, e.g., relevant in biomedical applications where the latent force corresponds to a chemical concentration, which can only be non-negative. 

In this work, we propose two optimal state estimation frameworks \citep{Rawlings2017} to perform physically consistent inference in nonlinear latent force models. First, we employ a continuous-time \textit{full information estimation} framework, which is particularly suitable for smaller datasets. Second, we introduce a continuous-time \textit{delayed moving horizon estimation} (MHE) framework, which can handle very large datasets. The frameworks are centered around optimization problems to determine an optimal state sequence that (i) follows the system dynamics, (ii) satisfies system-inherent constraints, and (iii) minimizes a cost function which takes the estimated disturbances into account. The consideration of system-inherent constraints is a distinct advantage of the proposed method since this is generally not possible for the nonlinear Kalman filter extensions. Moreover, optimal state estimation can outperform nonlinear Kalman filter extensions such as the unscented Kalman filter and the extended Kalman filter, compare the examples in \citep{Rawlings2017}. Finally, we evaluate the developed frameworks in a variety of examples. Among others, we consider a real-world biomedical system - the hypothalamic-pituitary-thyroid (HPT) axis \citep{Eisenberg2008,Dietrich2001,Wolff2022}. This example is particularly suitable for the application of latent force models as there exist thoroughly developed mathematical models and since the dynamics are driven by an unmeasurable and uncontrollable hormone called thyrotropin releasing hormone ($TRH$) that acts as latent force. 

The remaining of the article is structured as follows. In Section~\ref{sec:lfm}, we recapitulate latent force models and in Section~\ref{sec:mhe:scheme}, we introduce the novel optimal state estimation frameworks. We demonstrate the performance of the proposed frameworks in several numerical examples in Section~\ref{sec:numerical:evaluation} and conclude this work in Section~\ref{sec:conclusion}.

\textit{Notation:} We denote the set of real numbers by $\mathbb{R}$, the set of all integers greater than or equal to $a$ by $\mathbb{I}_{\geq a}$ and the identity matrix of dimension $N$ is denoted by $I_N$. For a vector $ x = \begin{bmatrix}
	x_1 & \dots & x_n
\end{bmatrix}^\top \in \mathbb{R}^n$ and a positive definite matrix $P$, the weighted vector norm is written as $||x||_P = \sqrt{x^\top P x}$. If $P=I$, we simply write $||x||$, denoting the Euclidean norm of the vector $x$. A diagonal matrix with entries $q_1, \dots, q_n$ is denoted by $\mathrm{diag}(q_1, \dots, q_n)$. 


\section{Latent force models}
\label{sec:lfm}
In this section, we start by recapitulating the state-space representation of GP priors \citep{Hartikainen2010}. Then, we explain how this state-space representation is exploited for inference in latent force models and we discuss the relation of latent force models and unknown input observers (UIOs). Finally, we motivate the use of optimal state estimation in latent force models by demonstrating that the state-of-the-art approaches can lead to an unconvincing performance.

\subsection{State-space representation of Gaussian processes}
\label{subsec:state:space:GP}
Latent force models heavily rely on the concept of GP regression which is a versatile Bayesian machine learning technique to model unknown (nonlinear) functions from which input/output data is available. A priori available function properties such as smoothness or periodicity can be encoded by choosing a particular covariance function. A detailed introduction to GPs can be found in \citep{Rasmussen2006}. GP priors can be represented by state-space models that are driven by white noise \citep{Hartikainen2010}, as we briefly explain in the following. We first consider a stochastic differential equation 
\begin{align}
	&\frac{d^{m_f} }{dt^{m_f}}f_{\mathrm{GP}}(t) + a_{{m_f}-1} \frac{d^{{m_f}-1} }{dt^{{m_f}-1}}f_{\mathrm{GP}}(t) + \nonumber\\
	& \hspace{2cm}\dots + a_1 \frac{d}{dt} f_{\mathrm{GP}} (t) + a_0f_{\mathrm{GP}}(t) = w_\mathrm{GP}(t) \label{sde:lfm}
\end{align}
with $f_{\mathrm{GP}}: \mathbb{R} \rightarrow \mathbb{R}$ being an ${m_f}$-times differentiable stochastic process with the stationary covariance function $k:\mathbb{R}\times \mathbb{R} \rightarrow \mathbb{R}$, known constants $a_0, a_1, \dots, a_{m_f-1} \in \mathbb{R}$ and $w_{\mathrm{GP}}$ being white noise with some spectral density $S_{w_{\mathrm{GP}}}(\omega) = q_{\mathrm{GP}}$ with $\omega$ being the frequency. The stochastic differential equation~(\ref{sde:lfm}) can be reformulated as
\begin{align}
	\label{def:GP:state:space}
	\dot{x}_{\mathrm{GP}}(t) = Fx_{\mathrm{GP}}(t) + Lw_\mathrm{GP} 
\end{align}
with $x_{\mathrm{GP}}(t) = \begin{pmatrix}
	f_\mathrm{GP}(t) & \frac{df_\mathrm{GP}(t)}{dt} & \dots & \frac{d^{{m_f}-1}f_\mathrm{GP}(t)}{dt^{{m_f}-1}}
\end{pmatrix}^\top$ and 
\begin{align}
	F = \begin{pmatrix}
		0 & 1 & \dots & 0 \\
		\vdots & \ddots & \ddots  & \vdots \\
		0 & \dots & 0 & 1 \\  
		-a_0 & \dots & -a_{{m_f}-2} & -a_{{m_f}-1} \\
	\end{pmatrix} \: L = \begin{pmatrix}
		0 \\ \vdots \\ 0 \\ 1
	\end{pmatrix}.
\end{align}
We can retrieve the original GP as $f_\mathrm{GP} = C_\mathrm{GP}x_\mathrm{GP}$ with $C_\mathrm{GP} = \begin{pmatrix}
	1 & 0 &\dots & 0
\end{pmatrix}$. In the following, it will be useful to extract the power spectral density of $f_\mathrm{GP}$ \citep{Hartikainen2010}
\begin{align*}
	S_{f_\mathrm{GP}} = C_\mathrm{GP} (F + j\omega I)^{-1}Lq_{\mathrm{GP}}L^\top \big((F-j\omega I)^{-1}\big)^\top C_{\mathrm{GP}}^\top.
\end{align*}
We exploit that the power spectral density $S_{f_\mathrm{GP}}$ and the (stationary) covariance function~$k$ constitute a Fourier pair \citep[p. 149]{Carlson2010}. Hence, if one wants to represent a GP with a specific covariance function by the state-space representation (\ref{def:GP:state:space}), we need to set the constants  $a_0, a_1, \dots, a_{{m_f}-1}$ and $ q_{\mathrm{GP}}$ such that the power spectral densities of the covariance function and the state-space representation match. This can be exactly done for Matérn covariance functions, but approximations are needed for the squared exponential covariance function \citep{Hartikainen2010}. Posterior distributions can be obtained by applying the well known Kalman filter, introduced by \cite{Kalman1960} and the RTS smoother, see \cite{Rauch1965}.

\subsection{Linear latent force models}
\label{subsec:latent:force:models}
In linear latent force models, we assume to have knowledge of our system dynamics such that we can establish
\begin{subequations}
	\label{def:linear:system:dynamics}	
	\begin{align}
		\dot{x}(t) &= Ax(t) + Bu(t) + G\ell(t) + Ew(t) \\
		y(t) &= Cx(t) + v(t)
	\end{align}
\end{subequations}
with states $x \in \mathbb{R}^n$, control inputs $u \in \mathbb{R}^{m_u}$, outputs $y \in \mathbb{R}^p$, latent force $\ell \in \mathbb{R}$, disturbances $w \in  \mathbb{R}^q$, $v\in \mathbb{R}^p$, and matrices $A, B, G, E, C$ of appropriate dimensions. As done in \citep{Hartikainen2011,Grasshoff2019}, we can straightforwardly augment the dynamics~(\ref{def:linear:system:dynamics}) with the state-space representation of GP priors~(\ref{def:GP:state:space}) (where~$\ell$ from~(\ref{def:linear:system:dynamics}) corresponds to the first state (i.e. $f_\mathrm{GP}$) of $x_\mathrm{GP}$ in (\ref{def:GP:state:space})) resulting in the \textit{latent force model}
\begin{subequations}
	\label{def:latent:force:model}
	\begin{align}
		\dot{x}_a(t) & = A_a x_a(t) + B_a u(t) + E_aw_a(t)\\
		y(t) &= C_a x_a(t) + v(t)
	\end{align}
\end{subequations}
with $x_a = \begin{pmatrix}
	x^\top  & x_{\mathrm{GP}}^\top
\end{pmatrix}^\top$, $w_a = \begin{pmatrix}
	w^\top & w_{\mathrm{GP}}^\top
\end{pmatrix}^\top$, and 
\begin{align*}
	A_a &= \begin{pmatrix}
		A & GC_{\mathrm{GP}}\\
		0 & F 
	\end{pmatrix}, \quad B_a = \begin{pmatrix}
		B \\ 0
	\end{pmatrix}, \quad E_a = \begin{pmatrix}
		E & 0 \\
		0 & L
	\end{pmatrix} \\
	C_a &= \begin{pmatrix}
		C & 0
	\end{pmatrix}.
\end{align*} 
To get posterior distributions of the latent force and of the states, one can apply the Kalman filter/RTS smoother (KF/RTSS) approach as suggested in \citep{Hartikainen2011}. The GP hyperparameters are computed by maximizing the marginal likelihood, which is a side-product of the Kalman filter recursions \citep{Hartikainen2011}. 
\begin{remark}
	\label{rmk:relation:UIO}
	Latent force models have similarities with the concept of UIOs. Both methods estimate the states of a system, which is subject to some unknown input (latent force). To obtain the estimates, many UIO approaches use Luenberger-like observers \citep{Chen2012}, whereas latent force models are often based on the application of the KF/RTSS approach \citep{Hartikainen2011,Grasshoff2019}. One key difference between both approaches is that latent force models model the unknown input as a GP. This can be beneficial as prior knowledge about the latent force (such as smoothness or periodicity) can be encoded in the covariance function. However, this representation may not respect inherent constraints of the latent force, such as, e.g., in case of a latent force that can only be non-negative. A second key difference is that UIOs aim for an online estimation of the states, whereas latent force models are commonly used to perform regression, which is also the focus of this work.
\end{remark}

\subsection{Motivating Example: Reconstruction of Transcription Factors}
\label{subsec:motivating:example}
\begin{figure}[t!]
	\centering
	\includegraphics[scale=0.5]{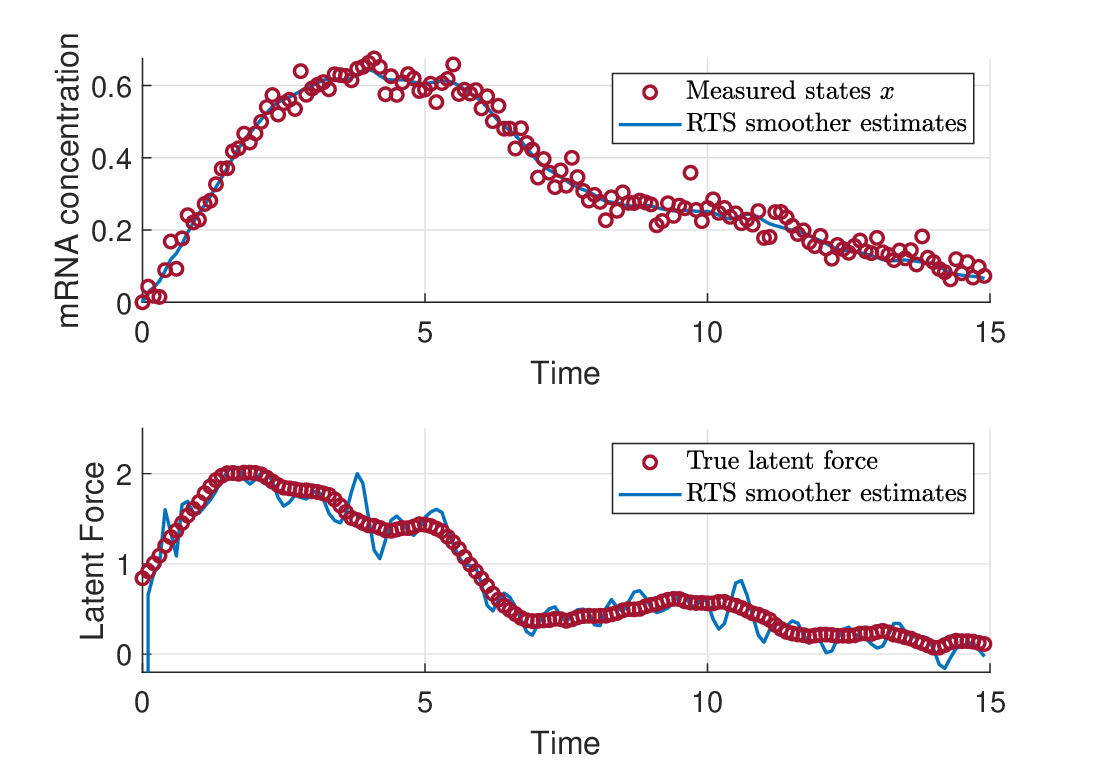}
	\caption{Simulation results of the KF/RTSS approach to obtain estimates of the state and the latent force for system~(\ref{ex:motivating:example}).}
	\label{fig:motivating:KF:RTS}
\end{figure}

In case of correctly determined hyperparameters and correctly specified noise variance, the KF/RTSS approach to perform inference for linear latent force models works accurately \citep{Hartikainen2011}. However, if this is not the case, for instance due to a small amount of training data, we observed in simulations that the performance of the KF/RTSS approach declines. This observation is illustrated in the following example. Consider the system
\begin{align}
	\label{ex:motivating:example}
	\dot{x}(t) = S\ell(t) - Dx(t) + w(t),
\end{align}
with the sensitivity $S = 0.25$ and the decay rate $D = 0.6 \frac{1}{\mathrm{s}}$ and some normally distributed process noise $w$ with zero mean and standard deviation $\sigma_w = 10^{-3}$.
In this system, which is considered in a similar fashion in \citep{Lawrence2006,Hartikainen2012}, the state $x$ corresponds to an mRNA concentration and the latent force $\ell$ to a transcription factor. We assume to have noisy state measurements available, where the measurement noise is normally distributed with standard deviation $\sigma_v = 0.025$. We consider $\ell(t) \sim \mathcal{GP}(m(t),k(t,t'))$ with a Matérn covariance function. As in~(\ref{def:latent:force:model}), we augment the dynamics~(\ref{ex:motivating:example}) with the state-space representation of the GP resulting in the following augmented state-space model 
\begin{align}
	\dot{x}_a(t) = \begin{pmatrix}
		-D & S & 0 \\
		0 & 0 & 1 \\
		0 & -\lambda^2 & -2\lambda \\
	\end{pmatrix}x_a(t) + \begin{pmatrix}
		1 & 0 \\
		0 & 0 \\
		0 & 1 \\
	\end{pmatrix} w_a(t)
\end{align}
with $\lambda = \sqrt{2\nu_m}/\ell_{\mathrm{ls}}$ and $\nu_m = p_m + 1/2$, $p_m =1$, and $\ell_{\mathrm{ls}}$ being the length scale of the Matérn covariance function. Note that a process modeled with a Matérn covariance function is $k_{m}$ times mean-square differentiable if and only if $\nu_m >k_m$ \citep[Ch.~4]{Rasmussen2006}. Then, we use the KF/RTSS approach to reconstruct the latent force. To this end, we discretize the system with a sampling time of $\delta = 0.01 \: \mathrm{s}$ and optimize the hyperparameter~$\ell_{\mathrm{ls}}$ during the first 60 samples and use the fixed hyperparameter afterwards. The initial estimate is set according to $\hat{x}(0) \sim \mathcal{N}(0, \Sigma_\infty)$ with $\Sigma_\infty$ as outlined in \citep[Sec. 2.2]{Hartikainen2011}. As shown in Figure~\ref{fig:motivating:KF:RTS}, the estimates of the latent force are improvable, due to some fluctuations. More importantly, the estimates of the latent force take some negative values, which is physically implausible since the latent force corresponds to a chemical concentration. This performance can be explained by the misspecified hyperparameters due to a rather short training time and by the fact that we cannot consider constraints in the estimation scheme.


\section{Optimal State Estimation Frameworks for Latent Force Models}
\label{sec:mhe:scheme}
In this section, we start with an introduction of the considered setting. Then, we introduce the novel optimal state estimation frameworks for inference in latent force models. 
\subsection{Setting}
\label{subsec:setting}
In the following, we consider general nonlinear systems with additive disturbances of the following form
\begin{subequations}
	\label{system:def}
	\begin{align}
		\dot{x}(t) &= f(x(t),u(t), \ell(t)) +w(t) \\
		y(t) &= h(x(t), u(t))+ v(t)
	\end{align}
\end{subequations}
with states $x\in \mathbb{R}^n$, control inputs $u \in \mathbb{R}^{m_u}$, latent force~$\ell \in \mathbb{R}$, disturbances $w \in \mathbb{R}^q$, $v \in \mathbb{R}^p$, outputs $y \in \mathbb{R}^p$, state transition function $f:\mathbb{R}^n \times \mathbb{R}^{m_u} \times \mathbb{R} \rightarrow \mathbb{R}^n$ and output map $h:\mathbb{R}^n \times \mathbb{R}^{m_u} \rightarrow \mathbb{R}^p$. Throughout this work, we assume that there exists a unique solution to~(\ref{system:def}). As done in \citep{Hartikainen2011,Hartikainen2012,Grasshoff2019,Landgraf2022}, we model the latent force as a GP as in (\ref{def:latent:force:model}) and augment the system dynamics~(\ref{system:def}) with its state-space representation and obtain
\begin{subequations}
	\label{system:def:augmented}
	\begin{align}
		\dot{x}_a(t) &= \begin{pmatrix}
			f(x(t),u(t), C_\mathrm{GP} x_\mathrm{GP}(t)) + w(t) \\ 
			F x_\mathrm{GP}(t) +Lw_{\mathrm{GP}}(t)
		\end{pmatrix} \label{system:dynamics:nonlinear:augmented}\\ 
		y(t) &= h_a(x_a(t), u(t))+ v(t) \label{system:output:nonlinear:augmented}
	\end{align}
\end{subequations}
with $x_a = \begin{pmatrix}
	x^\top & x_{\mathrm{GP}}^\top 
\end{pmatrix}^\top$ and $h_a: \mathbb{R}^{n+m_f} \times \mathbb{R}^{m_u} \rightarrow \mathbb{R}^p$. We assume to have prior knowledge (e.g., physically consistent estimates) in the form of $x_a \in \mathcal{X} \subseteq \mathbb{R}^{n+m_f}$, $u \in \mathcal{U} \subseteq \mathbb{R}^{m_u}$, $y \in \mathcal{Y} \subseteq \mathbb{R}^p$, $w_a  = \begin{pmatrix}
	w^\top & w_{\mathrm{GP}}^\top 
\end{pmatrix}^\top \in \mathcal{W} \subseteq \mathbb{R}^{n+m_f}$, and $v \in \mathcal{V}$. Given some initial state $\chi_a \in \mathcal{X}$, a control trajectory $u(t) \in \mathcal{U}$ for all $t\in \mathbb{\mathbb{R}}_{\geq 0}$, and a disturbance trajectory $w_a(t) \in \mathcal{W}$ for all $t \in \mathbb{\mathbb{R}}_{\geq 0}$, the solution to (\ref{system:def:augmented}) for all times $t \in \mathbb{R}_{\geq 0}$ is denoted by $x_a(t, \chi_a, u, w_a)$ and the corresponding output signal by $y(t) = h(x_a(t,\chi_a,u,w_a),u(t)) + v(t)$. 

\textit{Problem Statement:} Given the augmented nonlinear system~(\ref{system:def:augmented}) with an unknown latent force $\ell$ and output measurements $y$, we aim to reconstruct the latent force $\ell$ and the estimate the states~$x$.

In the here considered nonlinear case, we cannot perform exact inference anymore \citep{Hartikainen2012}. This is why approximate inference techniques must be used. To this end, we introduce two optimal state estimation frameworks to estimate the internal states (and thus also the latent force) of the augmented system~(\ref{system:def:augmented}) offline. Optimal state estimation is a suitable technique for this kind of problem due to its strong performance in practice as it has been shown that it can outperform the unscented Kalman filter and the extended Kalman filter, the robust stability guarantees, and the possibility to consider system-inherent constraints \citep{Rawlings2017,Schiller2023,Allan2021}. We note that the consideration of system-inherent constraints plays a particularly important role in the context of this work, compare Section~\ref{subsec:motivating:example}. If we model latent forces according to a GP, we implicitly allow it to become negative. This issue is known but cannot be solved by the KF/RTSS approach \citep{Hartikainen2012,Lawrence2006}. 

In the following, we propose two continuous-time optimal state estimation frameworks for inference in latent force models of the form (\ref{system:def:augmented}) in a regression context, i.e., in an offline setting, where all measurements have been collected before performing inference. In Section~\ref{subsec:full:optimization:problem}, we propose to use \textit{full information estimation} (FIE) \citep{Rawlings2017,Allan2021,Schiller2023} meaning that all available measurements are taken into account, which is beneficial for the performance and particularly useful in case we have small numbers of measurements available. In Section~\ref{subsec:mid:optimization:problem}, we introduce a delayed MHE scheme inspired by \citep{Schiller2025}, which can handle large numbers of measurements. 

\subsection{Full Information Estimation}
\label{subsec:full:optimization:problem}
Given some time length $T >0$ for which we have output measurements available, we solve the following optimization problem
\begin{subequations}
	\label{full:opt:problem}
	\begin{align}
		\min_{\bar{\chi}, \bar{w}}\quad  &J(\bar{\chi}, \bar{w}, \bar{y})\\
		\subject_to \bar{x}(\tau) &= x_a(\tau, \bar{\chi}, u, \bar{w}), \quad \tau \in [0, T] \label{full:constraints:dynamics}\\
		\bar{x} (\tau) &\in \mathcal{X}, \quad \tau \in [0,T] \label{full:state:constraints}\\
		\bar{y}(\tau) &= h_a(\bar{x}(\tau), u) + \bar{v}, \quad  \tau \in [0, T] \label{full:output:map} \\
		\bar{w}(\tau) &\in \mathcal{W}, \: \bar{y}(\tau) \in \mathcal{Y},\: \bar{v}(\tau) \in \mathcal{V}, \:  \tau \in [0,T]\label{full:output:disturbance:constraints}
	\end{align}
	with 
	\begin{align}
		J(\bar{\chi}, \bar{w}, &\bar{y}) = \Gamma(\bar{\chi},\hat{x}(0)) +\int_{0}^{T} L(\bar{w}(\tau), y(\tau) - \bar{y}(\tau))d\tau 
		\label{full:cost:function}
	\end{align}
	and (using $\Delta y = y - \bar{y}$)
	\begin{align}
		\Gamma(\chi,x) &= 2||\chi -x||_P^2 \label{full:cost:function:prior}\\
		L(w, \Delta y) &= 2||w||_Q^2 + ||\Delta y||_R^2.		\label{full:cost:function:details}
	\end{align}
\end{subequations}
The initial element of the estimated augmented signal is denoted by~$\bar{\chi}$ and the estimated noise signal by $\bar{w}:[0,T] \rightarrow \mathcal{W}$. In~(\ref{full:constraints:dynamics}),~$x_a$ denotes the solution to the system dynamics~(\ref{system:dynamics:nonlinear:augmented}) in dependence of the estimated initial state and the estimated process noise. With this constraint, we guarantee that the estimated states satisfy the system dynamics. Furthermore, by (\ref{full:state:constraints}) and (\ref{full:output:disturbance:constraints}), we constrain the estimated states, process and measurement disturbances, as well as the estimated outputs to satisfy the system-inherent constraints. In~(\ref{full:output:map}), we constrain the estimated output to follow the output map~(\ref{system:output:nonlinear:augmented}). In the cost function~(\ref{full:cost:function}) - (\ref{full:cost:function:details}), we consider stage costs which penalize the estimated noise and the difference of the measured outputs $y$ to the estimated outputs~$\bar{y}$ with some positive definite weighting matrices~$Q$ and~$R$ and a prior weighting~(\ref{full:cost:function:prior}) which penalizes the difference between the initial element of the estimated signal and the prior estimate $\hat{x}(0)$ with a positive definite weighting matrix~$P$. The solution to the problem directly gives us the estimated states and, thus, the estimates of the latent force $\hat{x}(t) = \bar{x}^\ast(t,\bar{\chi}^\ast, u,\bar{w}^\ast)$ for all $t \in [0,T]$. 

Although we consider a continuous-time setting, we require a sampling strategy, since we could not solve the optimization problem otherwise. Ultimately, we are interested in computing estimates of the true states at these sampling instants. This is standard in the literature on optimal control including model predictive control and MHE \citep{Rawlings2017,Schiller2024}.

The optimization problem (\ref{full:opt:problem}) uses the whole available dataset to determine the optimal state trajectory. Although using all output measurements is beneficial for the performance, solving (\ref{full:opt:problem}) might be unreasonably time-consuming for large datasets. In these cases, one might be interested in reducing the computational load. To this end, we introduce a second framework to estimate the states and the latent force, which is less computationally complex. 

\subsection{Delayed Moving Horizon Estimation}
\label{subsec:mid:optimization:problem}
One candidate to reduce the computational load is standard MHE. However, considering a delayed MHE scheme was shown to be beneficial compared to using a standard MHE scheme without delay \citep{Schiller2025}. In particular, the delay results in a performance of the obtained estimates that is close to the acausal infinite-horizon optimal estimator, compare also Section~\ref{sec:numerical:evaluation} for more details.

The idea is to solve repeatedly an optimization problem based on a moving horizon of length~$M << T$ (where $T$ denotes the time length for which we have measurements available). Once again, we require a sampling strategy. In this case, it will be convenient to define the set of all distinct time instances $t_i$ at which we take measurements (and at which we solve the optimization problem introduced below) by $\mathcal{T} \in \mathbb{R}_{\geq 0}$. We assume an equidistant sampling and that the horizon $M$ is an even integer multiple of the sampling time\footnote{In that case, the theoretical results from \cite{Schiller2025} can be transferred to the here considered continuous-time setting at the sampling instants, compare Section~\ref{sec:numerical:evaluation}.}~$\delta$.

The delayed MHE framework works similarly to standard MHE, with the difference that the estimated state is not set to the last element of the estimated state sequence, but to the middle element. In particular, we consider a moving time window $[t_{i} - M,t_i]$ and define the parts of the signals~$u$ and~$y$ within this time window as
\begin{align}
	u_{t_i}(\tau) &= u(t_i-M+\tau ) \qquad \tau \in [0, M) \label{def:u:window}\\
	y_{t_i}(\tau) &= y(t_i-M+\tau, \chi, u,w_a) \qquad \tau \in [0, M) \label{def:y:window}
\end{align}
with $\chi$ being the true (unknown) initial state. We repeatedly solve the following optimization problem for all $t_i \geq M$ to obtain the optimal estimated state trajectory within the time window $[t_{i} - M,t_i]$
\begin{subequations}
	\label{MHE:optimization:problem}
	\begin{align}
		\min_{\bar{\chi}_{t_i}, \bar{w}_{t_i}} &J(\bar{\chi}_{t_i}, \bar{w}_{t_i}, \bar{y}_{t_i}, t_i)\\
		\subject_to \bar{x}_{t_i}(\tau) &= x_a(\tau, \bar{\chi}_{t_i}, u_{t_i}, \bar{w}_{t_i}), \quad \tau \in [0, M] \label{MHE:constraints:first}\\
		\bar{x}_{t_i} (\tau) &\in \mathcal{X}, \quad \tau \in [0,M] \\
		\bar{y}_{t_i}(\tau) &= h_a(\bar{x}_{t_i}, u_{t_i}) + \bar{v}_{t_i}, \quad  \tau \in [0, M]\\
		\bar{w}_{t_i}(\tau) &\in \mathcal{W}, \quad \bar{y}_{t_i}(\tau) \in \mathcal{Y},\quad \bar{v}_{t_i}(\tau) \in \mathcal{V},\quad  \tau \in [0,M]\label{MHE:constraints:last}
	\end{align}
	with 
	\begin{align}
		J(\bar{\chi}_{t_i}, &\bar{w}_{t_i}, \bar{y}_{t_i}, t_i) = \Gamma(\bar{\chi}_{t_i},\hat{x}(t_i-M))\nonumber \\
		&+\int_{0}^{M} L(\bar{w}_{t_i}(\tau), y_{t_i}(\tau) - \bar{y}_{t_i}(\tau))d\tau 
		\label{MHE:cost:function}
	\end{align}
\end{subequations}
and $\Gamma$, $L$, as in (\ref{full:cost:function:details}), (\ref{full:cost:function:prior}), respectively, for some positive definite weighting matrices $R$, $Q$, $P$. The initial element of the estimated signal at time $t_i$ is denoted by $\bar{\chi}_{t_i}$ and~$\bar{w}_{t_i}:[0,T] \rightarrow \mathcal{W}$ denotes the estimated disturbance signal at time~$t_i$ (similarly for $\bar{v}_{t_i}$). The optimization problem (\ref{MHE:optimization:problem}) is similar to~(\ref{full:opt:problem}), except that we here consider a moving horizon and not the complete batch of measurements.

The solution to the optimization problem (\ref{MHE:optimization:problem}) at sampling time $t_i$ is denoted by $(\bar{\chi}_{t_i}^\ast, \bar{w}_{t_i}^\ast)$. The entire optimal estimated state signal is written as 
\begin{align}
	\label{eq:optimal:solution}
	\bar{x}_{t_i}^\ast(\tau) = x_a(\tau,\bar{\chi}_{t_i}^\ast, u_{t_i}, \bar{w}_{t_i}^\ast), \tau \in [0,M].
\end{align}
For time instances $t_i \in \mathcal{T}$, which satisfy $M \leq t_i \leq T$, we fix the state estimates at times $t_i - M/2$ to
\begin{align}
	\hat{x}(t_i-M/2) \coloneqq \bar{x}_{t_i}^\ast(M/2).
\end{align}
The complete estimated (piece-wise continuous) state signal $\hat{x}(t)$ is then defined as
\begin{align}
	\hat{x}(t) \coloneqq \begin{cases}
		\bar{x}^{\ast}_M(t), \quad t \in [0,M/2]\\
		\bar{x}_{\nu(t)+ M/2}^\ast (t-\nu(t) + M/2), t \in (M/2, T- M/2]\\
		\bar{x}_T^\ast(t), \quad t \in (T -M/2, T]
	\end{cases}
	\label{def:sol:opt:prob:dMHE}
\end{align}
with $\bar{x}^\ast$ from (\ref{eq:optimal:solution}) and $\nu(t)$ corresponding to
\begin{align}
	\label{def:nu}
	\nu(t) \coloneqq \min_{\nu\in \{\nu\in \mathcal{T}: \nu\geq t\}} \nu.
\end{align}

\begin{remark}
	\label{rmk:online estimation}
	As mentioned before, we here consider offline regression. However, in order to perform online state estimation, we could still use framework~(\ref{MHE:optimization:problem}) to perform delayed online state estimation. Alternatively, one could apply the full information estimation framework (\ref{full:opt:problem}) with~$T$ corresponding to the current online time~$t$ at the price of a growing optimization problem.
\end{remark}

\subsection{Hyperparameter optimization}
\label{subsec:hyper:opt}
In the context of GPs, one typically optimizes the hyperparameters by maximizing the marginal likelihood \citep{Rasmussen2006}. When applying (nonlinear extensions of) the KF/RTSS  to perform inference, the hyperparameters are still determined by maximizing the marginal likelihood, since it is a side-product of the Kalman filter recursions \citep{Hartikainen2011,Grasshoff2019}. However, in optimal state estimation, we do not inherently compute an expression related to the marginal likelihood. This is due to the conceptual differences between Kalman filtering and  nonlinear optimal state estimation. 

Here, we suggest to estimate the hyperparameters by doing standard parameter estimation. This means that we consider the parameters as states with constant dynamics and augment the state-space model once more \citep{Haykin2004}. In the estimation process, we then determine the optimal hyperparameters.

\section{Validation in numerical examples}
\label{sec:numerical:evaluation}
A theoretical analysis of the performance of the (discrete-time) optimal estimation framework~(\ref{MHE:optimization:problem})-(\ref{def:nu}), i.e., using a delayed MHE scheme, has been done in \citep{Schiller2025,Schiller2024}. The results were derived for a discrete-time setting. It is easy to show that they also hold at the sampling instants in the here considered continuous-time setting. The analysis is centered around the performance of an omniscient infinite-horizon estimator. In particular, using a delay in the MHE scheme is crucial to show that the obtained performance is close to the performance of the omniscient estimator given a suitable turnpike property \citep[Rmk. 8]{Schiller2025}. Furthermore, this implies accurate performance in the sense that the estimated states are close to the unknown true states \citep[Prop. 3]{Schiller2025}). In addition, it is established that the performance of the discrete-time version of the full information estimation framework~(\ref{full:opt:problem}) (without prior weighting) is close to the performance of the omniscient estimator \citep[Sec. V]{Schiller2025}.

In this work, we validate the performance of the proposed optimal state estimation frameworks for latent force models by applying them to three different examples. First, we consider the motivating example described in Section~\ref{subsec:motivating:example}, in which the KF/RTSS approach lacks accuracy and fails to result in physically meaningful estimates. Second, we consider a nonlinear ballistic target example from \cite{Hartikainen2012}. Finally, we consider a highly nonlinear and complex biomedical example, the HPT axis, compare, e.g., the works of \cite{Dietrich2001,Wolff2022,Eisenberg2008} for which we have real measurements from a clinical study available.

\subsection{Motivating Example: Reconstruction of Transcription Factors}
\label{subsec:MHE:motivating:example}
\begin{figure}[t!]
	\includegraphics[scale=0.5]{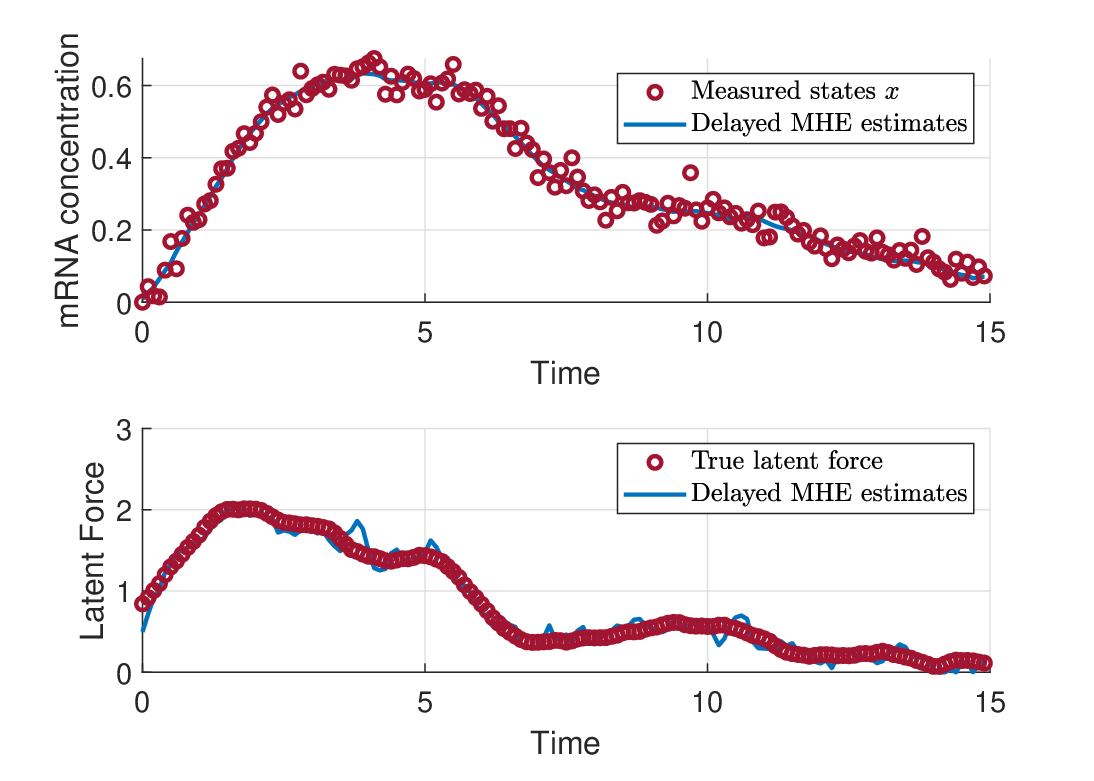}
	\centering 
	\caption{Simulation results of applying scheme~(\ref{MHE:optimization:problem}) to the system introduced in Section~(\ref{subsec:motivating:example}).}
	\label{fig:motivation:MHE}
\end{figure}

We consider the same example and the same setting, i.e., the same sampling time $\delta = 0.01 \: \mathrm{s}$, initial estimate $\hat{x}(0)$, and time interval to learn the hyperparameter\footnote{After the initial time interval, we fix the hyperparameter to the last estimate of the time interval.}, as in Section~\ref{subsec:motivating:example} and implement the scheme from~(\ref{MHE:optimization:problem}) since we have a large number of measurements available (in total 1500 samples). We consider a horizon length $M = 60$. In the cost function, the weighting matrices~$Q$ and~$R$ are set according to the inverse of the noise variances (compare Section~\ref{subsec:motivating:example}) and $P$ is chosen as $P = 10^{-2}I_n$ since we assume a poor initial guess. We constrain the estimates of the latent force to be non-negative. The estimation results can be seen in Figure~\ref{fig:motivation:MHE}. It becomes evident that the newly introduced framework outperforms the KF/RTSS approach, compare Figure~\ref{fig:motivating:KF:RTS}. The mean squared error (MSE) of the estimates of the latent force using framework (\ref{MHE:optimization:problem})-(\ref{def:nu}) after the initial time interval to learn the hyperparameters corresponds to 0.0103, whereas the MSE of the KF/RTSS approach corresponds to 0.0192, resulting in a 86 \% larger MSE. Moreover, the estimation framework does not result in physically implausible state estimates. 


\subsection{Ballistic Target}
\label{subsec:ballistic:target}
In this subsection, we analyze the performance of the scheme~(\ref{full:opt:problem}) introduced in Section~\ref{subsec:full:optimization:problem}  with an example from~\cite{Hartikainen2012} since we do not have many measurements available. It represents a ballistic target that approaches the earth surface in 1D. The first state represents the altitude and the second state the velocity of the target. In this example, we only have a nonlinear output map available. The exact dynamics are
\begin{align*}
	\dot{x}(t) &= \begin{pmatrix}
		-x_2(t) +w_1(t)\\
		-\alpha e^{-\gamma x_1(t)} x_2(t)^2 + g + \ell(t) + w_2(t)
	\end{pmatrix}\\
	y&= \sqrt{s_x^2 + (s_y -x_1(t))^2} + v(t)
\end{align*}
with $\alpha = 4.49 \cdot 10^{-4}$, $\gamma = 1.49 \cdot 10^{-4}$, $g = 9.81$, $s_x = 30000$, and $s_y =  30$. The acceleration is due to a drag force, gravity~$g$, and some unknown force~$\ell$, which corresponds to the latent force. 
\begin{figure}[t!]
	\includegraphics[scale=0.5]{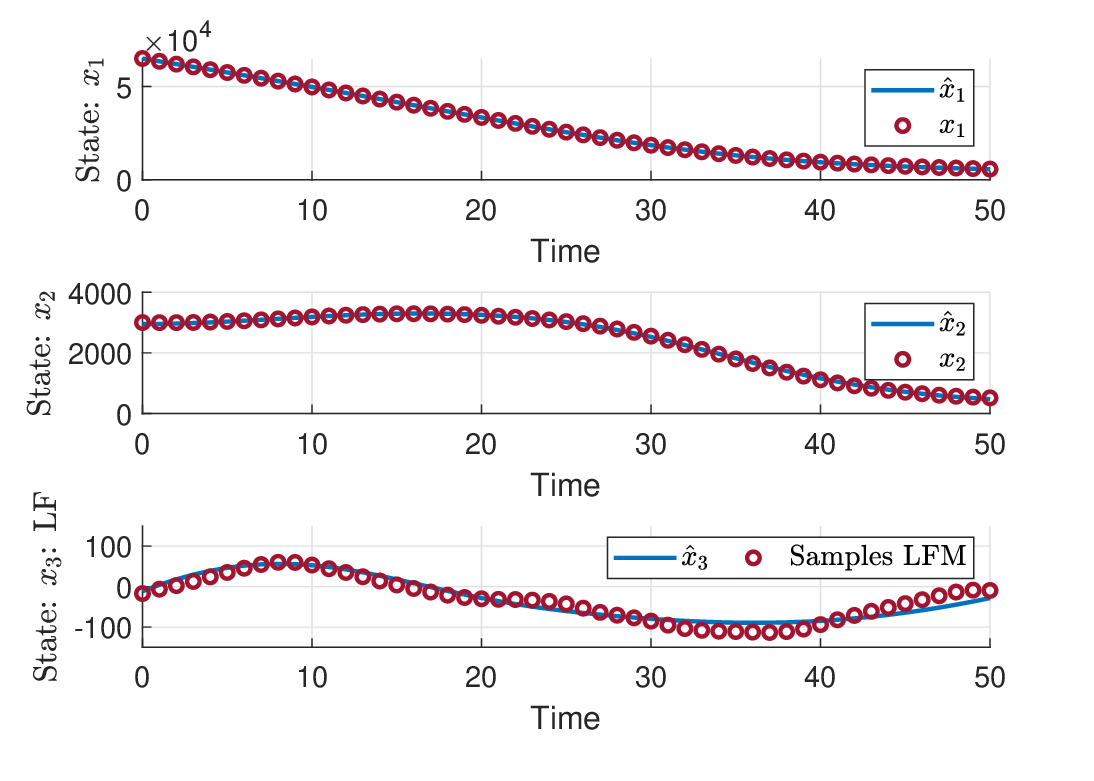}
	\centering 
	\caption{Simulation results of the scheme introduced in Section~\ref{subsec:full:optimization:problem} for system~(\ref{BT:augmented:model}). The abbreviations LF and LFM stand for ``latent force" and ``latent force model", respectively.}
	\label{fig:ballistic}
\end{figure}

As in \citep{Hartikainen2012}, we model the latent force as a GP with a Matérn covariance function and $p_m=2$ (which results in three additional states). Furthermore, we estimate the unknown hyperparameter (i.e., the length scale of the Matérn covariance function) by treating it as a state resulting ultimately in 6 states. Altogether, we obtain the following augmented state-space model
\begin{subequations}
	\label{BT:augmented:model}
	\begin{align}
		\dot{x}(t) &= \begin{pmatrix}
			-x_2(t) +w_1(t)\\
			-\alpha e^{-\gamma x_1(t)} x_2(t)^2 + g + x_3(t) + w_2(t)\\
			x_4(t) +w_3(t)\\
			x_5(t) +w_4(t)\\
			-\lambda^3 x_3(t) -3\lambda^2x_4(t) - 3\lambda x_5(t)+ w_5(t)\\
			w_6(t)
		\end{pmatrix}
	\end{align}
	with the output map
	\begin{align}
		y&= \sqrt{s_x^2 + (s_y -x_1(t))^2} + v(t).
	\end{align}
\end{subequations}
We simulate the system for 25 seconds with $x(0) = \begin{pmatrix}
	65000 & 3000
\end{pmatrix}^\top$ for the physical part (i.e., the first two states) of the system. We assume to be able to measure the state every $\delta = 0.5 \: \:\mathrm{s}$. We consider normally distributed process noise with mean zero and variance $\Sigma_w =  \mathrm{diag}(50, 10)$ and normally distributed measurement noise with mean $\mu_v = 0$ and variance $\sigma_v^2 = 900$, as in \citep{Hartikainen2012}. Then, we start the estimation process with the initial estimate $\hat{x}(0) = \begin{pmatrix}
	55000 & 2000
\end{pmatrix}^\top$. As weighting matrices, we consider $Q= \mathrm{diag}(10,10,1,1,10,1)$, $R = 1$, and $P =10^{-11}I_n$ (the small weight in the prior weighting is again due to the assumption of having a poor initial guess). The results of the implementation are illustrated in Figure~\ref{fig:ballistic}. As one can see from this figure, the latent force is well reconstructed although we cannot measure it. The states are also precisely estimated. 
We observe a slight degradation of the performance for the last samples. This observation may be due to the estimates of the FIE framework being on the \textit{leaving arc}, compare the work of \cite{Schiller2025}.

\subsection{Application to the Pituitary-Thyroid-Feedback Loop}
\label{subsec:hpt}
\begin{figure}[t!]
	\centering
	\vspace{5pt}
		\begin{tikzpicture}[scale=0.9]
		\draw[line width = .5mm, color=mygreen](3,0.25) -- (7,0.25);
		\draw[line width = .5mm, color=mygreen,->](7,0.25) -- (7,-0.9);
		\draw[line width = .5mm, color=myblue](3,-0.25) -- (3.525,-0.25);
		\draw[line width = .5mm, color=myblue,->](3.5,-1) -- (4,-1);
		\node[align=center] at (3.25,.5) {$T_3$};
		\draw[line width = .5mm, rounded corners] (1,-.5) rectangle (3,.5);
		\node[align=center] at (2,0) {Thyroid};
		\draw[line width = .5mm,->, color=mygreen](6,-1) -- (6.9,-1);
		\draw[line width = .5mm, rounded corners] (4,-1.5) rectangle (6,-.5);
		\node[align=center] at (5,-1) {D1 \& D2};
		\node[align=center] at (6.5,-1.25) {$T_3$};
		\draw[line width = .5mm,->, color=mygreen](7.1,-1) -- (7.5,-1);
		\draw[line width = .5mm] (7,-1) circle (0.1);
		\draw[line width = .3mm](6.9,-1) -- (7.1,-1);
		\draw[line width = .3mm](7,-0.9) -- (7,-1.1);
		\node[align=center] at (3.2,-1) {$T_4$};
		\draw[line width = .5mm, color=myblue](3.5,-0.25) -- (3.5,-2.025);
		\draw[line width = .5mm,->, color=myblue](3.5,-2) -- (3,-2);
		\draw[line width = .5mm, rounded corners] (1,-2.5) rectangle (3,-1.5);
		\node[align=center] at (2,-2) {Pituitary};
		\node[align=center] at (1.4,-1) {$TSH$};
		\draw[line width = .5mm,->, color=myred](2,-1.5) -- (2,-0.5);
		\draw[line width = .5mm, rounded corners] (0.5,-4.5) rectangle (3.5,-3.5);
		\node[align=center] at (2,-4) {Hypothalamus};
		\draw[line width = .5mm,->](2,-3.5) -- (2,-2.5);
		\node[align=center] at (1.4,-3) {$TRH$};
\end{tikzpicture}
	\caption{Block diagram of the HPT axis. }
	\label{fig:HPT_axis}
\end{figure}
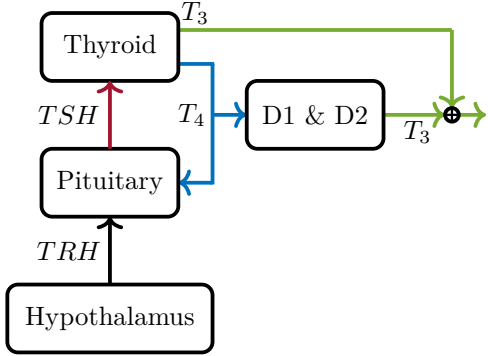

Next, we consider a considerably more complex real-world example, namely the HPT axis, which is a natural control loop in the human body \citep{Greenspan1997}. The mechanisms are illustrated in Figure~\ref{fig:HPT_axis}. The hormone $TRH$ stimulates at the pituitary the production of the thyroid stimulating hormone ($TSH$), which in turn stimulates the production of triiodothyronine ($T_3$) and thyroxine ($T_4$) at the thyroid gland. The hormone $T_4$ is converted into $T_3$ by the enzymes 5'-deiodinase type I (D1) and 5'-deiodinase type~II (D2). Moreover, $T_4$ inhibits the production of $TSH$ at the pituitary. In the following, we use a thoroughly developed nonlinear (6-dimensional) mathematical model of the HPT axis \citep{Dietrich2001,Wolff2022}. The dynamics and the related numerical parameter values of the HPT axis are given in \citep[Supp. Material]{Wolff2022}. 
$TRH$ cannot be controlled or measured, nor does it represent a disturbance. Hence, we model it as latent force with a Matérn kernel and $p_m=1$. As in the previous subsection, we estimate the length scale in the estimation process by performing a standard parameter estimation, where the parameter is modeled as state with zero dynamics.

Different to the previous example, we here have real hourly measurements of the hormone concentrations $TSH$, $FT_3$ (free version of $T_3$, simple proportional relation to $T_3$), and $FT_4$ (free version of $T_4$, again simple proportional relation to $T_4$) of~28 patients available for 24 hours \citep{Russell2008}. Due to significant noise levels in the individual measurements from \cite{Russell2008}, we average the hormone measurements at each time instant over the~28 patients, meaning that we draw conclusions for a generic person, not for the individual patients. Please note that no measurements of $TRH$ are available meaning that we have no information about the ground truth of this hormone concentration.

For this state-space model, we implement the scheme presented in Section~\ref{subsec:full:optimization:problem} to estimate the $TRH$ concentrations (once again because we do not have many measurements available). We use the following weighting matrices $R = 10 I_p$, $P = 10^{-8}I_n$, and $Q = \mathrm{diag}(10^2,10^2,10^2,10^2,10^4,10^2,5\cdot 10^{-5},5\cdot 10^{-5},1)$ and constrain the states to be non-negative. The states corresponding to the latent force have a significantly smaller weight which is due to the fact that we want to allow variations of $TRH$ to reconstruct the hormone concentrations.

\begin{figure}[t!]
	\includegraphics[scale=0.39]{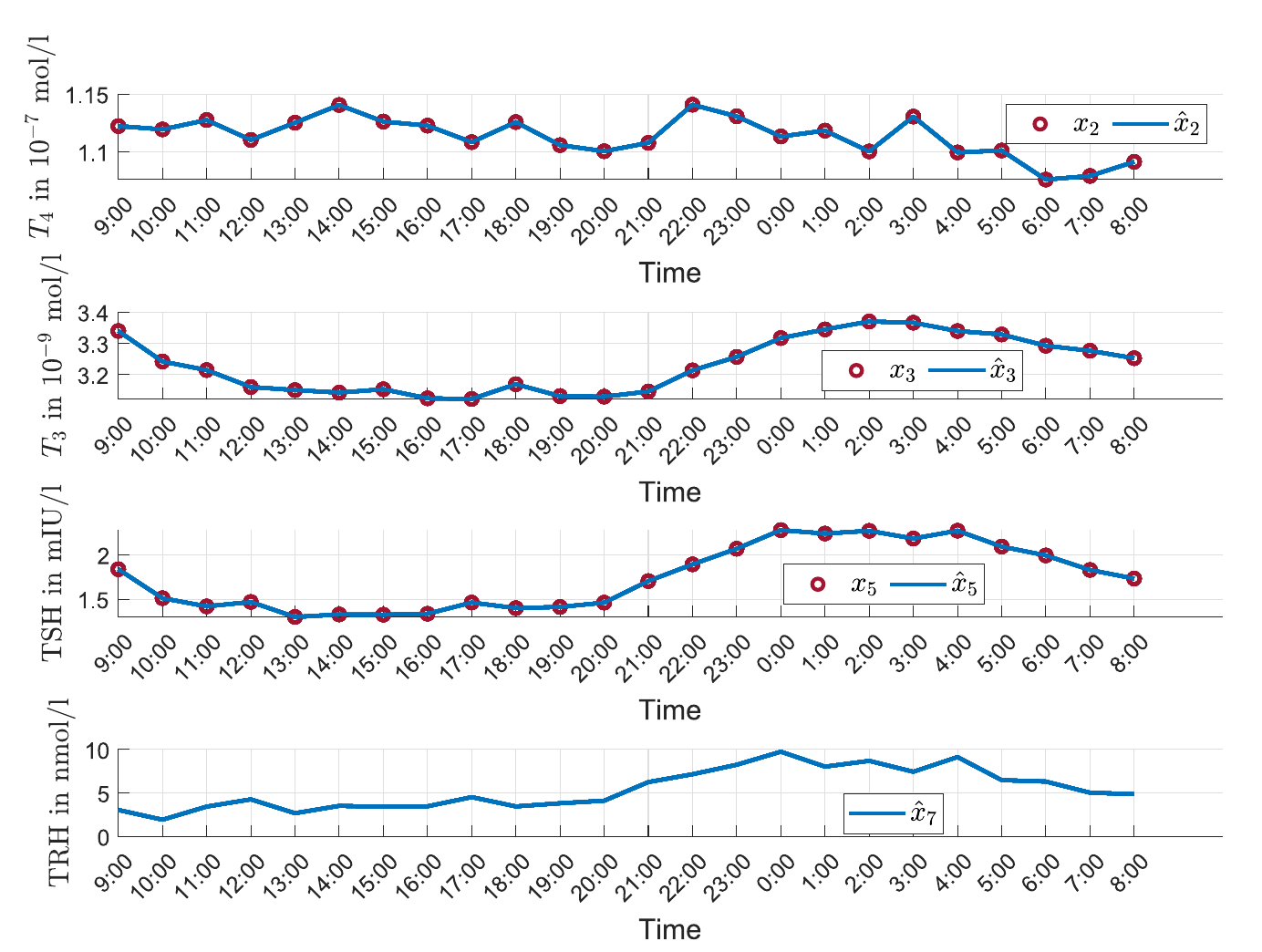}
	\centering 
	\caption{Simulation results for the reconstruction of the $TRH$ concentration in the HPT axis.}
	\label{fig:HPT}
\end{figure}

The simulation results are illustrated in Figure~\ref{fig:HPT}. As one can see from this figure, the measured states are very well reconstructed. Since the $TRH$ concentration cannot be measured, no comparison of the results to a ground truth is possible. However, the reconstructed daily $TRH$ concentration has a wide pulsatile shape, which is in line with results from clinical studies \citep{Brabant1991}. The reconstructed $TRH$ concentration could be beneficial to design individual optimal medication strategies \citep{Wolff2022}.

\section{CONCLUSION}
\label{sec:conclusion}
In this work, we introduced optimal state estimation frameworks to perform inference in latent force models. The \textit{full information estimation} framework exploits all available measurements to perform inference, whereas the \textit{delayed MHE} framework can handle large datasets. These frameworks can consider system-inherent constraints and outperformed state-of-the-art approaches to perform inference in latent force models in simulation studies. Furthermore, we demonstrated the potential of our approach by applying it to a highly nonlinear real-world example. 

An interesting subject for future work is an in-depth theoretical analysis of the detectability and observability properties of latent force (state-space) models. 

\section{Acknowledgment}
We wish to thank Prof. Richard Ross for generously providing data related to the HPT axis and Jan Graßhoff for sharing code related to the hyperparameter optimization in linear latent force models.

{\bibliography{ifacconf}}             
\end{document}